# Conceptual Knowledge Markup Language: The Central Core

Robert E. Kent
TOC (The Ontology Consortium)
550 Staley Dr.
Pullman, WA 99163, USA
rekent@ontologos.org

**ABSTRACT**

The conceptual knowledge framework OML/CKML needs several components for a successful design (Kent, 1999). One important, but previously overlooked, component is the central core of OML/CKML. The central core provides a theoretical link between the ontological specification in OML and the conceptual knowledge representation in CKML. This paper discusses the formal semantics and syntactic styles of the central core, and also the important role it plays in defining interoperability between OML/CKML, RDF/S and Ontolingua.

## OVERVIEW

The OML/CKML pair of languages is in various senses both description logic based and frame based. A bird's eye view of the architectural structure of OML/CKML is visualized in Figure 1.

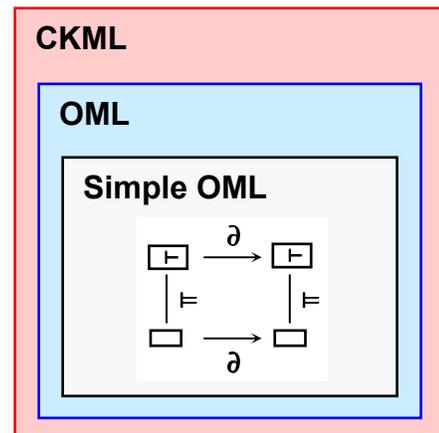

**Figure 1: OML/CKML at a glance**

- **CKML:** This language provides a conceptual knowledge framework for the representation of distributed information. Earlier versions of CKML followed rather exclusively the philosophy of Conceptual Knowledge Processing (CKP) (Wille, 1982; Ganter and Wille, 1989), a principled approach to knowledge representation and data analysis that "advocates methods and instruments of conceptual knowledge processing which support people in their rational thinking, judgment and acting and promote critical discussion." The new version of CKML continues to follow this approach, but also incorporates various principles, insights and techniques from Information Flow (IF), the logical design of distributed systems (Barwise and Seligman, 1997). This allows diverse communities of discourse to compare their own information structures, as coded in ontologies, logical theories and theory interpretations, with that of other communities that share a common terminology and semantics.

Beyond the elements of OML, CKML also includes the basic elements of information flow: classifications, infomorphisms, theories, interpretations, and local logics. The latter elements are discussed in detail in a future paper in preparation on the CKML knowledge model. Being based upon conceptual graphs, formal concept analysis, and information flow, CKML is closely related to a description logic based approach for modeling ontologies. Conceptual scaling and concept lattice algorithms correspond to subsumption.

- **OML:** This language represents ontological and schematic structure. Ontological structure includes classes, relationships, objects and constraints. How and how well a knowledge representation language expresses constraints is a very important issue. OML has three levels for constraint expression as illustrated in Figure 1.5:
    - top – sequents
    - intermediate – calculus of binary relations
    - bottom – logical expressions

  The top level models the theory constraints of information flow, the middle level arises both from the practical importance of binary relation constraints and the category theoretic orientation of the classification-projection semantics in the central core, and the bottom level corresponds to the conceptual graphs knowledge model with assertions (closed expressions) in exact correspondence with conceptual graphs.

  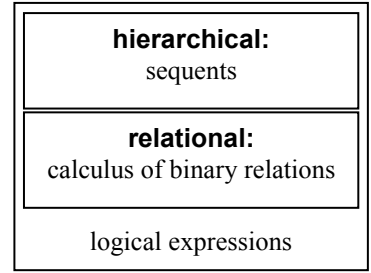

  **Figure 1.5: Constraints**

- **Simple OML:** This language is intended for interoperability. Simple OML was designed to provide the closest approach within OML to RDF/S, while still remaining in harmony with the underlying principles of CKML. In addition to the central core of CKML, Simple OML represents functions, reification, cardinality constraints, inverse relations, and collections. This paper shows how the first-order form of Simple OML is closely related to the Resource Description Framework with Schemas (RDF/S), and how the higher-order form of Simple OML is intimately related to XOL (XML-Based Ontology Exchange Language), an XML expression of Ontolingua with the knowledge model of Open Knowledge Base Connectivity (OKBC).

- **The Central Core:** This is based upon the fundamental classification-projection semantics illustrated in Figure 2. The expression of types and instances in the central core is very frame-like. In contrast to the practical bridge of the conceptual scaling process, the central core provides a theoretical bridge between OML and CKML.

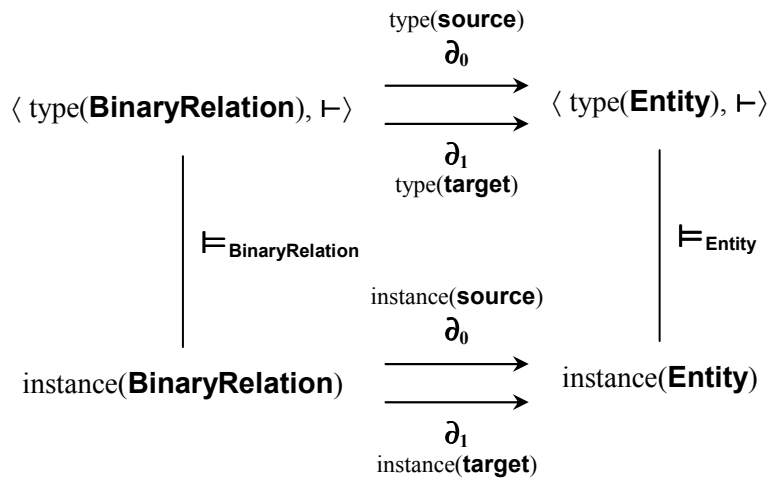

**Figure 2: Classification Projection Diagram**

# SEMANTICS

## Classification-Projection Diagram

In this section we define formal semantics for the fundamental classification-projection diagram illustrated by Figure 2. Figure 2 has two dimensions, the instance versus type distinction and the entity versus binary relation distinction. There are no subtype or disjointness constraints along either dimension. In Figure 2, arrows denote projection functions, lines denote classification relations, and type names denote higher order types (meta-types). Not visible in Figure 2 are the two entity types **Object** and **Data. Object** is the metaclass for all object types, whereas **Data** is the metaclass for all datatypes either primitive (such as strings, numbers, dates, etc.) or defined (such as enumerations). The **Entity** type is partitioned as a disjoint union or type sum, **Entity** = **Object** + **Data**, of the **Object** type and the **Data** type. So data values are on a par with object instances, although of course less complex.

The top subdiagram of Figure 2 owes much to category theory and type theory. A category is defined to be a collection of objects and a collection of morphisms (arrows), which are connected by two functions called source (domain) and target (codomain). To complete the picture, the composition and identity operators need to be added, along with suitable axioms. Also of interest are the various operators from the calculus of binary relations (Pratt, 1992), such as residuation. The partial orders on objects and arrows represent the type order on entities and binary relations. The bottom subdiagram gives a pointed version of category theory, a subject closely related to elementary topos theory. The classification relation connects the bottom subdiagram (instances) to the top subdiagram (types), and represents the classification relation of Barwise's Information Flow (Barwise and Seligman, 1997).

## Core Constraints

| symbol | meaning |
|---|---|
| $\rho : \alpha \to \beta$ | $\partial_0(\rho) = \alpha, \partial_1(\rho) = \beta$ |
| $\sigma : \gamma \to \delta$ | $\partial_0(\sigma) = \gamma, \partial_1(\sigma) = \delta$ |
| $r = (a, b)$ | $\partial_0(r) = a, \partial_1(r) = b$ |
| $r = \rho(a, b)$ | $\partial_0(r) = a, \partial_1(r) = b, r \vDash \rho$ |

**Table 1: Relational types**

Associated with the classification-projection diagram in Figure 2 are the following axiomatic properties. In the discussion below let $r$ be a relation instance having source entity $a$ and target entity $b$, let $\rho$ be a relation type having source type $\alpha$ and target type $\beta$, and let $\sigma$ be a relation type having source type $\gamma$ and target type $\delta$. This is symbolized in Table 1.

- **preservation of classification:**

    $r \vDash \rho$ **implies** ( $a \vDash \alpha$ **and** $b \vDash \beta$ )

    In words, if $r$ is an instance of (classified as) type $\rho$, then entity $a$ is an instance of type $\alpha$ and entity $b$ is an instance of type $\beta$. As an example, the citizenship relation is from the type Person to the type Country. If $c$ is an instance of citizenship, and $c$ relates $p$ to $n$, then $p$ is an instance of type Person and $n$ is an instance of type Country.

- **preservation of entailment:**

    $\sigma \vdash \rho$ **implies** ( $\gamma \vdash \alpha$ **and** $\delta \vdash \beta$ )

    The authorship binary relation from type Person to type Book is a subtype of the creatorship binary relation from type Agent to type Work. If a man *m* is an author of a book *b*, then the agent *m* is a creator of the work *b*. The facts that type Person is a subtype of type Agent and type Book is a subtype of type Work may be necessary conditions for the subtype relation.

- **inclusion implies subtype:**

    $\sigma \leq \rho$ **implies** $\sigma \vdash \rho$

    The motherhood binary relation on the type Person is a subtype of the parenthood binary relation on the type Person. If the woman *w* is the mother of a boy *b*, then *w* is a parent of *b*.

- **creation of incompatible types:**

    ( $\alpha, \gamma \vdash$ **or** $\beta, \delta \vdash$ ) **implies** $\rho, \sigma \vdash$

    The sibling relation on type Person is disjoint from the employment relation from type Person to type Organization. This is implied by the fact that type Person is disjoint from type Organization. This seems to be true in general, both for the source and target projections.

- **creation of incoherent type:**

    ( $\alpha \vdash$ **or** $\beta \vdash$) **implies** $\rho \vdash$

    If a relation type is specified to have a source (or target) entity type that is later found to be incoherent, then the relation type is also incoherent.

**Core Type Hierarchy**

The elaboration of the classification-projection diagram as depicted in Figure 3 illustrates the concepts (basic types) in the central core knowledge model. This model renders more explicitly the connections found in the Core Grammar. As a rule of thumb, XML elements become entity types in the core knowledge model, and attributes and content nonterminals (child embeddings) of XML elements become functions and binary relations. In Figure 3 a type is depicted by a rectangle and an instance is depicted by a bullet. The generic classification and subtype hierarchies have not been included as types (rectangles), since their instances are not needed until the full CKML is specified. When more than one subrectangle (subtype) is present, the subtypes partition the supertype. Instances of core relations and functions are listed and grouped within their appropriate types. The signatures and constraints for the core binary relations and functions are listed in Table 2.

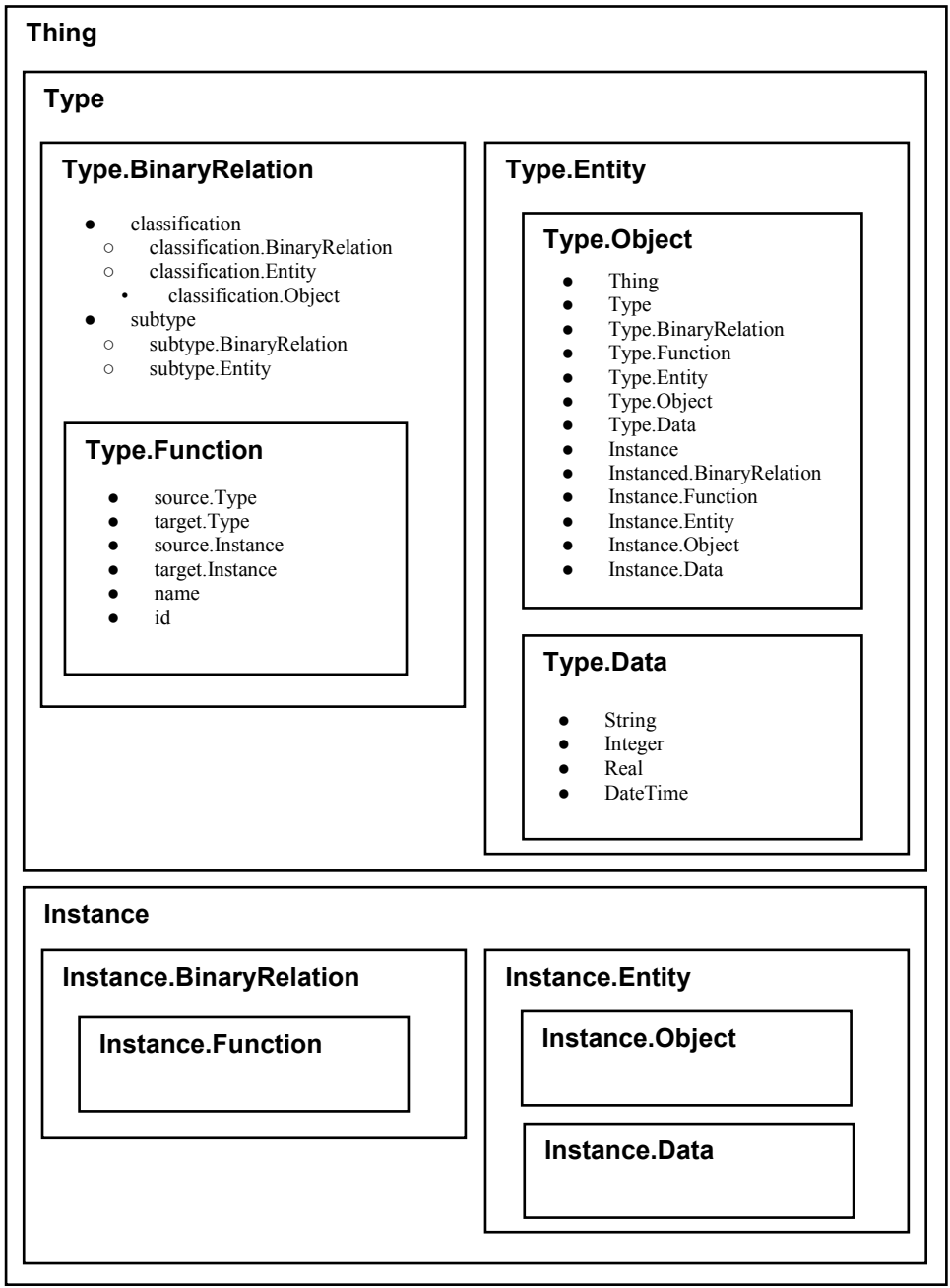

**Figure 3: Core Type Hierarchy**

### Binary Relations

```
classification : Instance → Type
        = classification.BinaryRelation + classification.Entity
classification.BinaryRelation : Instance.BinaryRelation → Type.BinaryRelation
classification.Entity : Instance.Entity → Type.Entity
        = classification.Object + classification.Data
classification.Object : Instance.Object → Type.Object
subtype : Type → Type
        =  subtype.BinaryRelation + subtype.Entity
subtype.BinaryRelation : Type.BinaryRelation → Type.BinaryRelation
```

```
subtype.Entity : Type.Entity → Type.Entity
comment : Thing → String
```

**Functions**

```
source.Type : Type.BinaryRelation → Type.Entity
target.Type : Type.BinaryRelation → Type.Entity
source.Instance : Instance.BinaryRelation → Instance.Entity
target.Instance : Instance.BinaryRelation → Instance.Entity
name : Type → String
id : Instance → String
```

Table 2: Core Signatures and Constraints

**Extended Operations**

A *graph*, as in Figure 3.5, is a set *N* of nodes, a set *E* of edges, and two functions called *source* $\partial_0$ and *target* $\partial_1$. In a graph the set of composable pairs of edges is the set

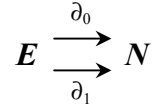

**Figure 3.5: Graph**

$$E\times_N E = \{(\rho,\sigma) \mid \rho,\sigma \in E \text{ and } \partial_1(\rho) = \partial_0(\sigma)\}.$$

Replacing nodes with objects *O* and edges with arrows *A*, a *category* is a graph with two additional functions

$$\iota : O \to A : A \mapsto \iota_A \qquad \circ : A\times_O A \to A : (\rho,\sigma) \mapsto \rho\circ\sigma$$

called *identity* and *composition*, satisfying the constraints

$$\partial_0(\iota_A) = A = \partial_1(\iota_A)$$

$$\partial_0(\rho\circ\sigma) = \partial_0(\rho) \text{ and } \partial_1(\rho\circ\sigma) = \partial_1(\sigma)$$

$$(\rho\circ\sigma)\circ\tau = \rho\circ(\sigma\circ\tau)$$

$$\iota_A\circ\rho = \rho \text{ and } \rho\circ\iota_B = \rho \text{ when } \partial_0(\rho) = A \text{ and } \partial_1(\rho) = B.$$

An *involution* in category is an function

$$(-)^\dagger : A \to A : \rho \mapsto \rho^\dagger$$

that satisfies the following constraints

$$\rho^{\dagger\dagger} = \rho^\dagger$$

$$\iota_A^\dagger = \iota_A$$

$$\partial_0(\rho^\dagger) = \partial_1(\rho) \text{ and } \partial_1(\rho^\dagger) = \partial_0(\rho)$$

$$(\rho\circ\sigma)^\dagger = \sigma^\dagger\circ\rho^\dagger$$

In OML/CKML the extended operations are as follows.

**Functions**

```
composition : Type.BinaryRelation × Type.BinaryRelation → Type.BinaryRelation
identity : Type.Entity → Type.BinaryRelation
transpose : Type.BinaryRelation → Type.BinaryRelation
```

The axiomatics for the subtype and classification core binary relations can be given either using the basics or using the composition and identity operators. The latter method is rather self-reflexive. The first axiom below states that the subtype relation is reflexive; more specifically, the identity relation is included in the subtype relation. The second axiom states that the subtype relation is transitive; more specifically, the composition of the subtype relation with itself is contained in the subtype relation. The third axiom states that the classificastion relation respects the subtype order; more specifically, the composition of the classificastion relation with the subtype relation is contained in the classification relation.

**axioms using the basics**

```
/* subtype reflexive */
<Forall var="t" type="Type">
  <subtype specific="t" generic="t"/>
</Forall>
/* subtype transitive */
<Forall var="t1 type="Type">
<Forall var="t2 type="Type">
<Forall var="t3 type="Type">
  <implies>
    <and>
      <subtype specific="t1" generic="t2"/>
      <subtype specific="t2" generic="t3"/>
    </and>
    <subtype specific="t1" generic="t3"/>
  </implies>
</Forall>
</Forall>
</Forall>
/* classification closure */
<Forall var="i" type="Instance">
<Forall var="t1" type="Type">
<Forall var="t2" type="Type">
  <implies>
    <and>
      <classification instance="i" type="t1"/>
      <subtype specific="t1" generic="t2"/>
    </and>
    <classification instance="i" type="t2"/>
  </implies>
</Forall>
</Forall>
</Forall>
```

**axioms using operators**

```
/* subtype reflexive */
<subtype specific="identity" generic="subtype"/>
/* subtype transitive */
<Forall var="r" type="BinaryRelation">
  <implies>
    <composition type="r" first="subtype" second="subtype"/>
    <subtype specific="r" generic="subtype"/>
  </implies>
</Forall>
/* classification closure */
<Forall var="r" type="BinaryRelation">
  <implies>
    <composition type="r" first="classification" second="subtype"/>
    <subtype specific="r" generic="classificastion"/>
  </implies>
</Forall>
```

## Core Grammar

Below we list a grammar for the central core that is relation-centric on types and object-centric on instances. Except for the inclusion of function types and instances, this grammar closely models the classification-projection diagram in Figure 2.

### oml bracket rule

```
 [1] oml             ::= '<OML>' ontology | collection '</OML>'
```

### ontology type rules

```
 [2] ontology    ::= '<Ontology>' (ext | typ | axm)* '</Ontology>'
 [3] ext         ::= '<extends' ontologyAttr prefixAttr '/>'
 [4] typ         ::= objType | binrelType | fnType
 [5] objType     ::= '<Type.Object' declTypeAttr '/>'
 [6] binrelType  ::= '<Type.BinaryRelation' declTypeAttr srcTypeAttr tgtTypeAttr '/>'
 [7] fnType      ::= `<Type.Function' declTypeAttr srcTypeAttr tgtTypeAttr '/>'
 [8] axm         ::= '<subtype' specificAttr genericAttr? '/>'
```

### collection instance rules

```
 [9] collection   ::= '<Collection' idAttr? ontologyAttr? '>' inst* '</Collection>'
[10] inst         ::= objInst
[11] objInst      ::= '<Instance.Object' idAttr? aboutAttr? '>'
                        (classInst | binrelInst | fnInst)*
                      '</Instance.Object>'
[12] binrelInst   ::= '<Instance.BinaryRelation' tgtInstAttr '>'
                         classInst*
                      '</Instance.BinaryRelation>'
[13] fnInst       ::= '<Instance.Function' tgtInstAttr '>'
                         classInst*
                      '</Instance.Function>'
[14] classInst    ::= '<classification' typAttr '/>'
```

### attribute rules

```
[15] ontologyAttr   ::=        'ontology = "' URI-reference '"'
[16] prefixAttr     ::=          'prefix = "' name '"'
[17] declTypeAttr   ::=            'name = "' name '"'
[18] srcTypeAttr    ::=     'source.Type = "' typeNSname '"'
[19] tgtTypeAttr    ::=     'target.Type = "' typeNSname '"'
[20] specificAttr   ::=        'specific = "' typeNSname '"'
[21] genericAttr    ::=         'generic = "' typeNSname '"'
[22] typAttr        ::=            'type = "' typeNSname '"'
[23] tgtInstAttr    ::= 'target.Instance = "' instanceNSname '"'
[24] idAttr         ::=              'id = "' name '"'
[25] aboutAttr      ::=           'about = "' URI-reference '"'
```

### basic XML rules

```
[26] typeNSname     ::= [ name ':' ] name
[27] instanceNSname ::= [ typeNSname '#' ] name
[28] URI-reference  ::= string, interpreted per [URI]
[29] name           ::= (any legal XML name symbol)
[30] string         ::= (any XML text, with "<", ">", and "&" escaped)
```

As indicated in the XML specification document an attribute name must be of the following form. In particular, the '.' is appropriate inside attribute names.

```
NameChar ::= Letter | Digit | '.' | '-' | '_' | ':' | CombiningChar | Extender
Name     ::= (Letter | '_' | ':') (NameChar)*
```

**Core DTD**

The elements, attributes and entities in the Core DTD below are tightly connected with the nonterminals and rules of the Core Grammar. The type elements are relation-centric (with respect to the `subtype` relation), whereas the instance elements are object-centric (with respect to the `classification` relation). The parameter entities `OML:Type`, `OML:Axiom` and `OML:Instance` represent in the DTD the "things" in the Core Type Hierarchy and Classification-Projection Diagram that are not represented by an XML tag.

**Parameter Entity Declarations**

```
<!-- rule [4] of the grammar -->
<!ENTITY % OML:Type
        "(OML:Type.Object
        | OML:Type.BinaryRelation
        | OML:Type.Function)">

<!-- rule [8] of the grammar -->
<!ENTITY % OML:Axiom
        "(OML:subtype)">

<!-- rule [10] of the grammar -->
<!ENTITY % OML:Instance
        "(OML:Instance.Object)">
```

**Element Type Declarations**

*oml bracket element*

```
<!-- rule [1] of the grammar -->
<!ELEMENT OML:OML (OML:Ontology | OML:Collection)>
```

*central core ontology dtd*

```
<!-- rule [2] of the grammar -->
<!ELEMENT OML:Ontology (OML:Extends | &OML:Type; | &OML:Axiom;)*>

<!-- rules [3], [15], [16] of the grammar -->
<!ELEMENT OML:extends EMPTY>
<!ATTLIST OML:extends
        ontology        CDATA #REQUIRED
        prefix          CDATA #IMPLIED>

<!-- rules [5], [17] of the grammar -->
<!ELEMENT OML:Type.Object EMPTY>
<!ATTLIST OML:Type.Object
        name            CDATA #REQUIRED>

<!-- rules [6], [17], [18], [19] of the grammar -->
<!ELEMENT OML:Type.BinaryRelation EMPTY>
<!ATTLIST OML:Type.BinaryRelation
        name            CDATA #REQUIRED
        source.Type     CDATA #REQUIRED
        target.Type     CDATA #REQUIRED>

<!-- rules [7], [17], [18], [19] of the grammar -->
<!ELEMENT OML:Type.Function EMPTY>
<!ATTLIST OML:Type.Function
        name            CDATA #REQUIRED
        source.Type     CDATA #REQUIRED
        target.Type     CDATA #REQUIRED>

<!-- rules [8], [20], [21] of the grammar -->
<!ELEMENT OML:subtype EMPTY>
<!ATTLIST OML:subtype
```

```
        specific        CDATA #REQUIRED
        generic         CDATA #IMPLIED>
```

*central core collection dtd*

```
<!-- rule [9], [24], [15] of the grammar -->
<!ELEMENT OML:Collection (&OML:Instance;)*>
<!ATTLIST OML:Collection
        id              CDATA #IMPLIED
        ontology        CDATA #IMPLIED>

<!-- rules [11], [24], [25] of the grammar -->
<!ELEMENT OML:Instance.Object
  (OML:classification | OML:Instance.BinaryRelation | OML:Instance.Function)*
>
<!ATTLIST OML:Instance.Object
        id              CDATA #IMPLIED
        about           CDATA #IMPLIED>

<!-- rules [12], [22], [23] of the grammar -->
<!ELEMENT OML:Instance.BinaryRelation (OML:classification)*>
<!ATTLIST OML:Instance.BinaryRelation
        target.Instance CDATA #REQUIRED>

<!-- rules [13], [22], [23] of the grammar -->
<!ELEMENT OML:Instance.Function (OML:classification)*>
<!ATTLIST OML:Instance.Function
        target.Instance CDATA #REQUIRED>

<!-- rules [14], [22] of the grammar -->
<!ELEMENT OML:classification EMPTY>
<!ATTLIST OML:classification
        type            CDATA #REQUIRED>
```

**Higher-Order Entity Types**

A first-order ontology is an ontology without higher-order types. In a first-order ontology the notions of instances and individuals coincide. Higher-order types are types that have other types as their instances. This means that instances can be either individuals or types. Individuals are instances that are not types. With higher-order types the classification relation extends to types on its source, and the source and target projection functions for individual relations also extended to types. Color is an example of a second-order type

**Color** = { Red, Orange, Yellow, Green, Blue, Indigo, Violet }

which has first-order color types, such as Red, as instances. The conceptual graph in Figure 4, an example from (Sowa, 1999), represents the English phrase *a red ball*. Here the characteristic relation (**chrc**) links the concept of a ball to the concept of the red color [**Color:** Red] whose type label is the second-order type **Color** and whose referent is the first-order type Red. The conceptual graph maps to the following logical formula.

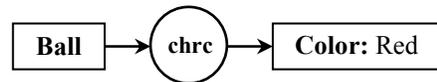

**Figure 4: higher-order type example**

($\exists$x:Ball)(color(Red) $\wedge$ chrc(x,Red)).

In the central core this can be represented as follows.

```
<Ontology>
    •••
  <Type.Object name="Color"/>
  <Type.Object name="Red"/>
    •••
```

```
    <classification instance="Red" type="Color"/>
       ...
    <Type.Object name="Ball"/>
    <Type.BinaryRelation name="chrc" source.Type="Ball" target.Type="Color"/>
</Ontology>

/* specific style */
<Collection>
       ...
    <Ball>
      <chrc target.Instance="Red"/>
    </Ball>
       ...
</Collection>
```

There are three things that are new here. An instance of the classification relation has been placed inside an ontology. The instance attribute of this classification refers to a type. The target attribute of the individual characteristic relation refers to a type.

We may also be interested in representing various relationships between types. For example, an "argument" relation (own slot) is from an object type to a multivalent relation type having that object as one of its arguments. In particular, the "Cast" ternary relation type in a Movie ontology has the "Movie" object type as one of its arguments.

```
<Ontology>
       ...
    <Type.BinaryRelation name="argument"
       source.Type="Type.Object" target.Type="Type.Relation"/>
       ...
    /* specific style */
    <argument source.Instance="Movie" target.Instance="Cast"/>
       ...
</Ontology>
```

There is one thing that is new here. An instance of the argument relation has been placed inside an ontology. Both the source and target attributes refer to types.

Figure 4 indicates how to extend the first-order classification-projection diagram of Figure 2 to higher-order entity types. As in the first-order case of Figure 2, the instance(**BinaryRelation**) metatype is the same as individual(**BinaryRelation**). However, the instance(**Entity**) metatype has changed to the sum **Entity** metatype, since object instances can be either individuals or types. The **Entity** metatype, representing entity instances, is the type sum (disjoint union) of its type and individual parts.

$$\textbf{Entity} = \text{type}(\textbf{Entity}) + \text{individual}(\textbf{Entity})$$
$$\text{instance}(\textbf{BinaryRelation}) = \text{individual}(\textbf{BinaryRelation})$$

The entity classification relation has been extended to include types at its source. This means that we can classify types with other higher-order types, ad infinitem. The source and target of individual binary relations have also been extended to include types. Note that the individual(**BinaryRelation**) metatype, along with its projection functions, correspond to frame-based own slots, whereas the type(**BinaryRelation**) metatype, along with its projection functions, correspond to frame-based template slots (see the With Ontolingua subsection below).

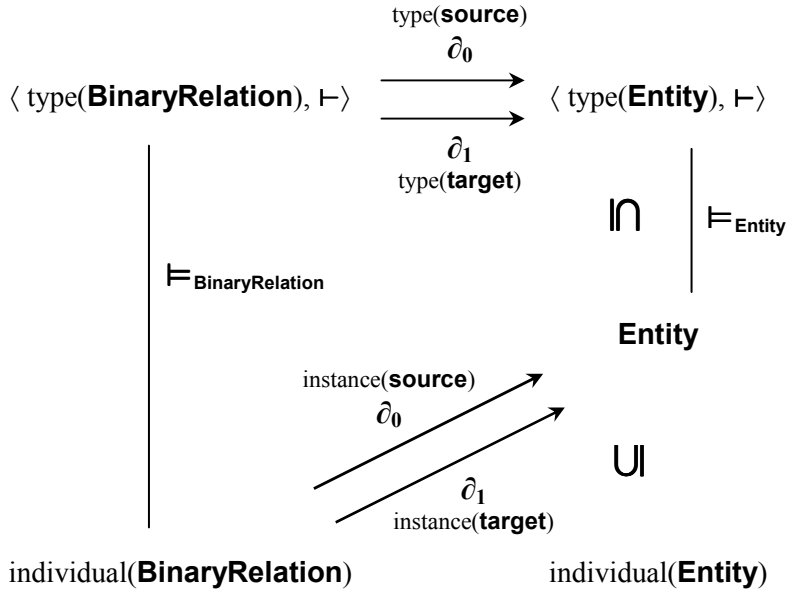

**Figure 4: Classification-Projection Diagram: Higher-Order Entity Types**

### Higher-Order Relation Types

Figure 5 displays the classification-projection diagam for higher-order types, not only for entities but also for relations. This is a further extension of, and very similar to, the first-order classification-projection diagram of Figure 2. Here the instance(**BinaryRelation**) metatype has changed to the sum **BinaryRelation** metatype, since relation instances can be either individuals or types. Since the **BinaryRelation** metatype is a type sum, the source and target functions are defined as copairings with the following definitions.

$$\textbf{source} \quad = \quad [\text{ type}(\textbf{source}) \circ \text{incl, individual}(\textbf{source}) \,]$$
$$\textbf{target} \quad = \quad [\text{ type}(\textbf{target}) \circ \text{incl, individual}(\textbf{target}) \,]$$

In addition, some explanation should be given for the definition of the classification relation for binary relations, that has now been lifted to types. This relation is the copairing of the following two binary relations.

$$\vDash_{\textbf{BinaryRelation}} \;:\; \text{type}(\textbf{BinaryRelation}) \rightarrow \text{type}(\textbf{BinaryRelation})$$

$$\vDash_{\textbf{BinaryRelation}} \;:\; \text{individual}(\textbf{BinaryRelation}) \rightarrow \text{type}(\textbf{BinaryRelation})$$

The first classification relation between relational types is new. The second is the usual first-order classification relation, where we identify individuals with instances (in that case).

One possible axiom for higher-order relation classification is the following.

- **preservation of classification:**

    $\sigma \vDash \rho$ **implies** $(\,\gamma \vDash \alpha \;\textbf{and}\; \delta \vDash \beta\,)$

Suppose that relational type σ is an instance of relational type ρ. If σ has source type γ and target type δ and ρ has source type α and target type β, then γ is an instance of α and δ is an instance of β. As an example how this might occur, let entity types α and β be any two second level types, and define a second-level binary relation ρ between α and β to be those first-level binary relations between first-level entity type instances of α and β.

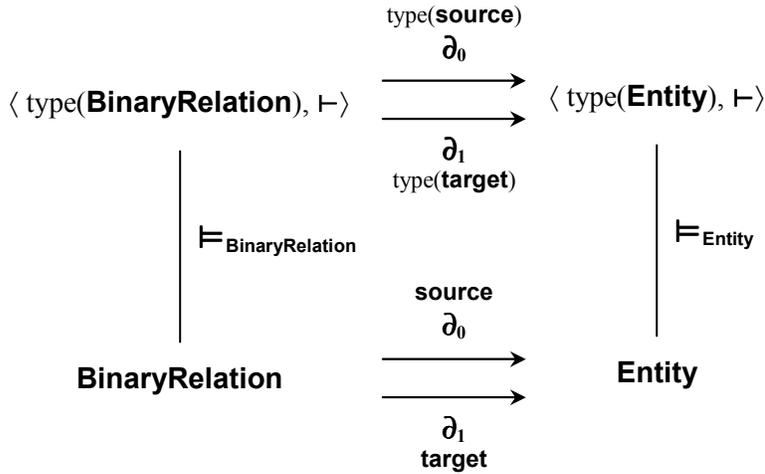

**Figure 5: Classification Projection Diagram: Higher-Ordered Types**

## SERIALIZATION SYNTAX

The National Center for Supercomputing Applications (NCSA) uses a search tool called Emerge that links multiple databases for a specialized community. Each community uses its own specialized markup language (XML application) for interchange of their particular information; for example, the astronomy community uses a special Astronomical Markup Language (AML). On the other hand, OML/CKML is a generic framework for describing information of any kind. What is the difference between a specialized markup language such as AML and a generic markup language (or framework) such as OML/CKML and how are these related? The answer involves coding and parsing styles.

The generic markup language XOL (see the section on interoperability) advocates a generic approach for the specification of ontologies. The generic approach means that all ontologically-structured information is specified by a single set of XOL tags (defined by the single XOL DTD). The generic approach is modeled in OML/CKML by the generic style discussed below. In contrast, the Conceptual Graph Interchange Form (CGIF) represents information in a specific style. The primary advantage for the generic approach is simplicity in language processing. The primary disadvantage is lack of a means for type-checking the semantic constraints specified in the ontology. As discussed in this section, OML/CKML offers an approach that subsumes both the generic and the specific approaches for coding ontologies and ontologically-structured information. In a nutshell, we want to investigate whether the equivalence of Figure 6 has any meaning, validity and importance. In fact, we believe it has central importance in processing ontologies and XML.

> **Ontology ≡ DTD**
>
> Figure 6: Equivalence

**Abbreviation Styles**

OML/CKML abbreviation styles are equivalent formalizations that have either the advantage of simpler processing (generic style) or the advantages of greater code simplicity and better type-checking (specific style). They are closely tied to the OML/CKML parsing methodology. There are two primary abbreviation styles: generic and specific. Any other style might be termed intermediate. The generic and specific styles are polar opposites, while an intermediate style is a mixture of the two. The generic style (no abbreviation) provides a syntax for a single universal grammar or DTD that is independent of domain and ontology. Each specific OML/CKML ontology can be automatically translated into a specific domain-dependent grammar or DTD. The specific style (full abbreviation) is an instance of that domain-specific ontology, and is parseable with that domain-specific grammar or DTD.

The OML/CKML abbreviation styles are based upon the two OML/CKML abbreviation forms; an object-element form and a function-attribute form. These loosely follow two of the three [RDF abbreviation forms](#) – the object-element form is essentially the third RDF abbreviation form with the RDF `Description` element corresponding to the OML/CKML `Instance.Object` element; the function-attribute form is essentially the first RDF abbreviation form restricted to OML/CKML functions. The object-element abbreviation form in OML/CKML preceded the RDF version by several years, providing the syntax for OML/CKML version 1.5. The generic style must use neither of these abbreviations, whereas the specific style must use both of them.

In order to illustrate OML/CKML abbreviation styles, we consider the example of the Movie instance Casablanca (1942). In the reduced representation below there is an object type for movies with metadata for year of appearance and genre. There is also a multivalent ($n$-ary) relation that links movies, cast members and the character that they played. The central core does not have a separate metatype for these (that comes in full OML), and so these are reified and represented as objects. The full Movie ontology can be automatically translated to the domain-specific movie DTD. Obviously, the specific style for Movie instance collections is much simpler code than the generic style.

**Movie Ontology**

```
<Type.Entity name="Movie"/>
<Type.Function name="year" source.Type="Movie" target.Type="Natno"/>
<Type.BinaryRelation name="genre" source.Type="Movie" target.Type="Genre"/>

<Type.Entity name="Cast"/>
<Type.Function name="movie" source.Type="Cast" target.Type="Movie"/>
<Type.Function name="member" source.Type="Cast" target.Type="Person"/>
<Type.Function name="character" source.Type="Cast" target.Type="String"/>
```

**Domain-Specific Movie DTD**

```
<!ELEMENT Movie (genre)*>
<!ATTLIST Movie
     id                    ID       #REQUIRED
```

```
        year                    NUMBER   #IMPLIED>

<!ELEMENT genre EMPTY>
<!ATTLIST genre
        target.Instance         CDATA    #REQUIRED>

<!ELEMENT Cast EMPTY>
<!ATTLIST Cast
        movie                   CDATA    #IMPLIED
        member                  CDATA    #IMPLIED
        character               CDATA    #IMPLIED>
```

**The Specific Style Collection**

```
<Movie id="Casablanca_1942" year="1942">
  <genre target.Instance="Drama"/>
  <genre target.Instance="Romance"/>
</Movie>

<Cast
   movie="Casablanca_1942"
   member="Humphrey_Bogart"
   character="Rich Blaine"/>
```

**The Generic Style Collection**

```
<Instance.Entity id="Casablanca_1942">
  <classification type="Movie"/>
  <Instance.Function target.Instance="1942">
    <classification type="year"/>
  </Instance.Function>
  <Instance.BinaryRelation target.Instance="Drama">
    <classification type="genre"/>
  </Instance.BinaryRelation>
  <Instance.BinaryRelation target.Instance="Romance">
    <classification type="genre"/>
  </Instance.BinaryRelation>
</Instance.Entity>

<Instance.Entity id="cast1">
  <classification type="Cast"/>
  <Instance.Function target.Instance="Casablanca_1942">
    <classification type="movie"/>
  </Instance.Function>
  <Instance.Function target.Instance="Humphrey_Bogart">
    <classification type="member"/>
  </Instance.Function>
  <Instance.Function target.Instance="Rich Blaine">
    <classification type="character"/>
  </Instance.Function>
</Instance.Entity>
```

The XML tags for both the ontology and the generic style instance collection use the generic names for types and instances in the central Core Type Hierarchy of Figure 3. These are listed in Table 3. The `subtype` and `classification` relations are special. The `subtype` relation needs the two additional `specific` and `generic` attributes, and the `classification` relation (since it links instances and types) needs the two additional `instance` and `type` attributes.

| central core type | generic kind | XML use |
|---|---|---|
| Type.BinaryRelation | object | tag |
| Type.Function | object | tag |
| `Type.Entity` | object | tag |
| `subtype` | binary relation | tag |
| `name` | binary relation | attribute |
| `source.Type` | binary relation | attribute |
| `target.Type` | binary relation | attribute |
| `Instance.BinaryRelation` | object | tag |
| `Instance.Function` | object | tag |
| `Instance.Entity` | object | tag |
| `classification` | binary relation | tag |
| `id` | binary relation | attribute |
| `source.Instance` | binary relation | attribute |
| `target.Instance` | binary relation | attribute |

**Table 3: The central core names for types and instances**

**Parsing**

Translation software can be developed that realizes the equivalence of Figure 6. There are two translational directions. The translational direction from DTDs to ontologies is exemplified by the Biopolymer ontology that was manually created from the intuitive semantics for the specific markup language BIOML, but not directly from its DTD. This direction is not intended to be an automatic translation, but instead requires domain expertise. Other examples such as this exist. The translational direction from ontologies to DTDs *is* straight-forward and automatic. Translation software can also be developed that translates between generic and specific style instance collections, using suitable collection DTDs. The processes involved in all of these translations are graphically illustrated in Figure 7. We discuss the first process in detail, but give the other two only a cursory glance.

*Ontology to Domain-Specific DTD Translation*

This is indicated as process **[1]** in Figure 7. Since all abbreviation styles and forms apply to instances only, the representation for an ontology is independent of the abbreviation styles. Since an ontology specified using the central core of OML/CKML must not use abbreviations, it *must* only use the generic type tags in Table 3. As a result, such an ontology can be automatically translated to a domain-specific DTD. The ontology serialization can be parsed with the central core ontology grammar or DTD, creating an internal representation for the ontology. The translation works on this internal ontology representation, producing a domain-specific DTD. The rules for translating from the internal representation for an OML/CKML ontology to a domain-specific DTD are as follows. This addresses one half of the equivalence in Figure 6. To follow this, use the Movie ontology as an example.

- Objects (entities) are represented as XML elements (tags).
  - Objects have *element content*. The content model consists of a repeatable choice of the binary relation elements that have the object as their first argument.
  - There is a required `id` attribute.
- Functions are represented as XML attributes.
  - Functions, as XML attributes, are all *implied*, since functions are partial and the central core does not have cardinality constraints (these occur first in Simple OML).
- Binary relations are represented as XML elements (tags).

- Binary relations have *empty content*.
- There is a required `target.Instance` attribute.

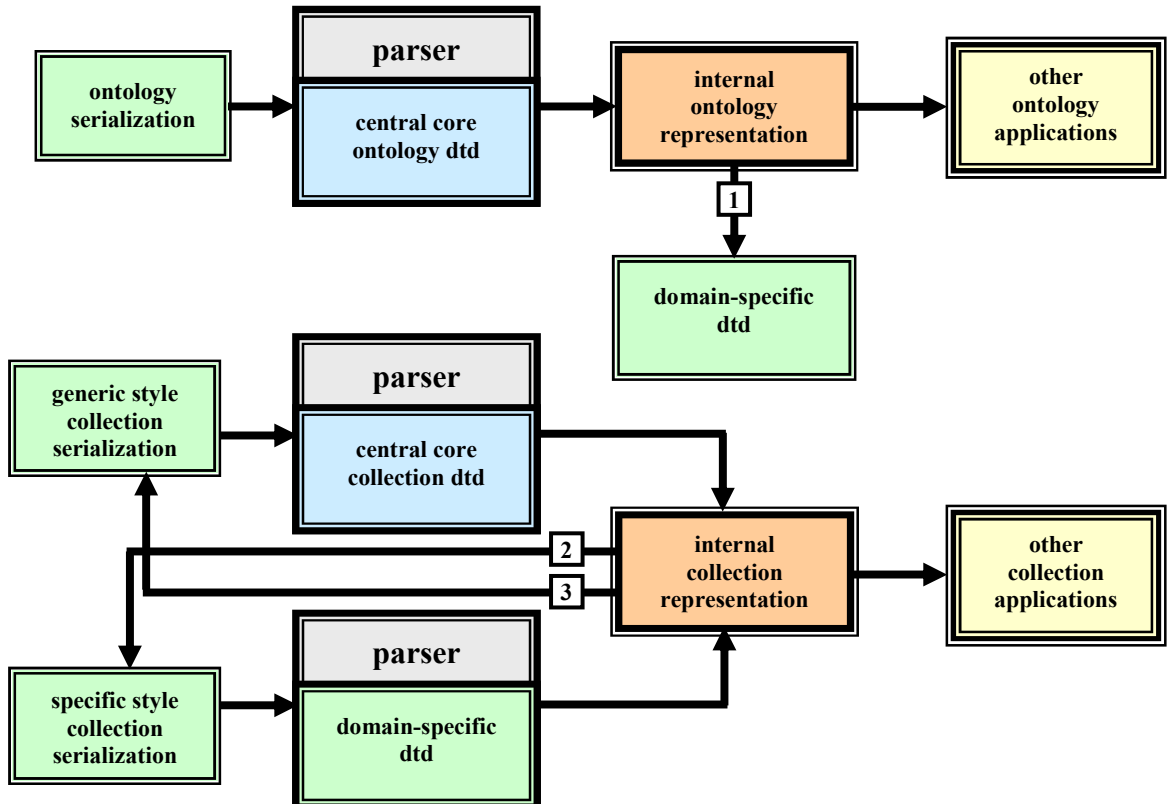

**Figure 7: Parsing Styles**

## Generic to Specific Instance Collection Translation

This is indicated as process **[2]** in Figure 7. To reiterate, abbreviation styles only apply to instance collections. The generic style collection serialization can be parsed with the central core collection grammar or DTD, creating an internal representation for the collection. The translation works on this internal collection representation, producing a specific style collection serialization. The specific style is characterized by the fact that *all* tags are non-generic, specific tags; that is, that *none* come from the central core instance names listed in Table 3. Also, *all* functions should be abbreviated as attributes.

## Specific to Generic Instance Collection Translation

This is indicated as process **[3]** in Figure 7. The specific style collection serialization can be parsed with the domain-specific DTD obtained from the first process **[1]**, creating an internal representation for the collection. The translation works on this internal collection representation, producing a generic style collection serialization. The generic style is characterized by the fact that *all* tags come from the central core instance names listed in Table 3. The function-attribute abbreviation is inoperative here.

**Higher-Order Entity Types**

In order to allow for the specification of higher-order entity types in the central core, the following changes must be made to the Core Grammar. Corresponding changes must also be made to the Core DTD.

1. Change the instance notation to individual.

2. Introduce **Entity**, the type sum of **Type.Entity** and **Individual.Entity**.

3. Allow classification instances to be specified in an ontology. This requires addition to the axiom production rule, and introduction of a new rule for instance attributes.

   **ontology type rules**

   ```
   axm ::= '<subtype' specificAttr genericAttr? '/>'
         | '<classification' instAttr typeAttr '/>'
   
   instAttr ::= 'instance = "' typeNSname '"'
   ```

4. In individuals change the target instance metatype from **Instance.Entity** to **Entity**. To accomplish this, do not change the target instance attribute to individual, but leave as instance. In addition, introduce an instance namespace name rule.

   **attribute rules**

   ```
   tgtInstAttr    ::= 'target.Instance = "' instanceNSname '"'
   instanceNSname ::= typeNSname | individualNSname
   ```

5. An instance of a binary relation between types corresponds to the frame-based notion of an own slot in a class. This can be handled by adding further to the axiom rule.

   **ontology type rules**

   ```
   axm ::= '<subtype' specificAttr genericAttr? '/>'
         | '<classification' instAttr typeAttr '/>'
         | '<Instance.BinaryRelation' typAttr srcTypeAttr tgtTypeAttr '/>'
   ```

**INTEROPERABILITY**

Interoperability is very important for a language whose goal is to represent distributed information in a conceptual framework. The discussion in this section demonstrates how CKML is interoperable with two important frame-based systems: Resource Description Framework with Schemas (RDF/S), and XOL, the XML expression of Ontolingua. Each of these is discussed in the following subsections.

**With RDF/S**

RDF/Schemas has the structure of a semantic network. It corresponds to simple conceptual graphs (Sowa, 1999), which are conceptual graphs without negations, universal quantifiers and nested conceptual contexts. The first-order classification-projection diagram in Figure 2 corresponds to RDF with type specification capabilities (RDF with Schemas). Elements of this correspondence are listed in Table 4. The question mark in Table 4 reflects the current undeveloped state of RDF/S data types. These are being developed by the XML Schema working group of the W3C, and will be incorporated into CKML when finalized.

| RDF/S notion | central core notion | central core formalism |
|---|---|---|
| Class | object type | type(**Object**) |
| ???? | data type | type(**Data**) |
| Property | binary relation type | type(**BinaryRelation**) |
| subClassOf | subtype on objects | $\vdash_{\mathbf{Entity}}$ |
| subPropertyOf | subtype on binary relations | $\vdash_{\mathbf{BinaryRelation}}$ |
| domain | type source | type(**source**) = $\partial_0$ |
| range | type target | type(**target**) = $\partial_1$ |
| Resource | object instance | instance(**Object**) |
| Literal | data type value | instance(**Data**) |
| Statement | binary relation instance | instance(**BinaryRelation**) |
| subject | instance source | instance(**source**) = $\partial_0$ |
| object | instance target | instance(**target**) = $\partial_1$ |
| predicate, type | classification | $\vDash_{\mathbf{BinaryRelation}}$, $\vDash_{\mathbf{Entity}}$ |

**Table 4: RDF/S and Simple OML Correspondences**

The fact that the first-order central core corresponds closely to the core structure of RDF/S (RDF/S without collections), illustrates why the core part of the RDF/S syntax is embeddable into the Simple OML syntax. The Simple OML serialization syntax is the closest approach to the RDF/S serialization syntax. The most obvious difference is the lack of types in basic RDF — these are to be modeled with schemas. Types are not considered as essential in RDF as they are in OML/CKML, since schema classes are just special kinds of RDF resources. This is reasonable and is close to the frame system approach, but it is different from the conceptual framework of OML/CKML, which is based on the theory of information flow (Barwise and Seligman, 1997). Although RDF Schema classes are normally modeled as types, in order to model the RDF semantics that "properties are resources," they could be modeled in OML/CKML as special objects, with explicit models for the subclass partial order relation between classes, the classification relation between resources and classes, the domain and range functions, etc.

There are several points at which the knowledge models for RDF/S and the OML/CKML central core differ.
1. In RDF/S everything is regarded to be a resource. So our correspondence between the RDF/S **Resource** metatype and the central core instance(**Object**) metatype is not accurate. A better solution would be to split the **Resource** metatype into two parts, so that it will correspond to the top level central core **Thing** metatype, in addition to the instance(**Object**) metatype.
2. In RDF/S the **Property** metatype, which corresponds to the central core type(**BinaryRelation**) metatype, is asserted to be a subtype of the **Resource** metatype. This is in agreement with the correspondence between the RDF/S **Resource** metatype and the central core **Thing** metatype, since in the OML/CKML central core the type(**BinaryRelation**) metatype is a subtype of the **Thing** metatype. However, it is not in agreement with the correspondence between the RDF/S **Resource** metatype and the central core instance(**Object**) metatype for two reasons:

(1) **Property** is a type notion, whereas instance(**Object**) is an instance notion; and (2) **Property** is a relation notion, whereas instance(**Object**) is an entity notion. In both category theory (and abstract graph theory) the set of objects (nodes) and the set of arrows (edges) have no constraints between them. This is the same idea that we asserted before: the dimension of the entity versus binary relation distinction in the fundamental classification-projection diagram of Figure 2 has no subtype or disjointness constraint. We also do not want to place any constraint on instances and type, especially for higher types as discussed below. This is the same idea that we asserted before: the dimension of the instance versus type distinction in the fundamental classification-projection diagram of Figure 2 has no subtype or disjointness constraint.

3. The correspondence between the RDF/S **Statement** metatype and the central core instance(**BinaryRelation**) metatype is not accurate. In RDF/S a statement is a triple of the form $(p, s, o)$, where $p$ is a property, $s$ is a resource, and $o$ is either a resource or a literal. Using the terminology in Table 1, we choose to interpret binary relation instances as pairs $r = (a, b)$ and not triples $r = \rho(a, b)$. Such a triple is an instance of a binary relation classification $(a, b) \models_{\text{BinaryRelation}} \rho$ between a binary relation instance $(a, b)$ and a binary relation type $\rho$. So the most accurate correspondence is the following.

| RDF/S notion | central core notion |
|---|---|
| Statement $(p, s, o)$ | binary relation classification $(a, b) \models \rho$ |

4. The OML namespace mechanism is a bit different from the RDF namespace mechanism. Any real-world object is represented by an OML object (surrogate) with a link to the real-world object and OML references to the real-world object are made through this surrogate, whereas web resources may be referenced in RDF without being described (represented). The complete references for an OML object (instance) has the 3-fold syntax *ontology*:*type*#*identifier*, an extension of the XML namespace mechanism.

**With Ontolingua**

XOL (XML Ontology Exchange Language) is a frame-based language with an XML syntax that is currently being designed for the exchange of ontologies for molecular biology. XOL produces an XML expression for Ontolingua through the OKBC application programming interface (API). In this section we show how the frame-based language XOL can be modeled by the central core of OML/CKML with higher-order entity types, the version of the classification-projection diagram as illustrated in Figure 4.

Figure 8 illustrates the type hierarchy for XOL. This corresponds to the core type hierarchy of Figure 3. The XOL types in Figure 8 originate in four ways. The three types **class**, **slot** and **individual** are the standard frame

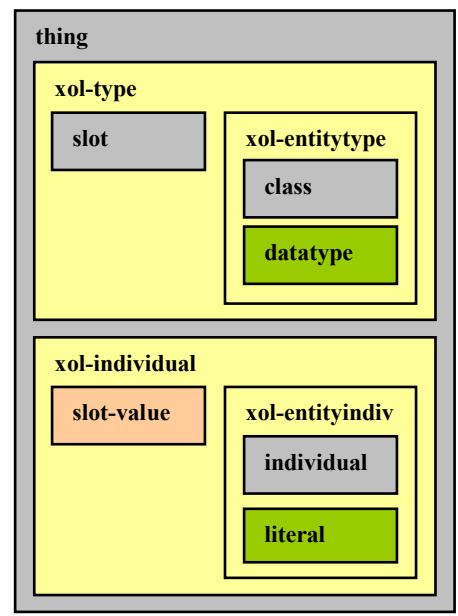

**Figure 8: XOL Type Hierarchy**

types. The type **thing** is the OKBC root type. The two types **datatype** and **literal** have been added for completeness. The type **slot-value** is a reified type. And, the four types

| | | |
|---:|:---:|:---|
| xol-type | = | slot + xol-entitytype |
| xol-entitytype | = | class + datatype |
| xol-individual | = | slot-value + xol-entityindiv |
| xol-entityindiv | = | individual + literal |

have been defined in order to organize the other types.

Here is the core aspect of the XOL DTD.

```
<!ELEMENT module
  (name, class*, slot*, individual*)
>
<!ELEMENT name (#PCDATA)>
<!ELEMENT class
  (name, (subclass-of | instance-of | slot-values)* )
>
<!ELEMENT slot
  (name, (domain | slot-value-type | slot-values)* )
>
<!ELEMENT individual
  (name, (instance-of | slot-values)* )
>
<!ELEMENT slot-values
  (name, value*)
>
<!ELEMENT subclass-of     (#PCDATA)>
<!ELEMENT instance-of     (#PCDATA)>
<!ELEMENT domain          (#PCDATA)>
<!ELEMENT slot-value-type (#PCDATA)>
```

From this DTD we can abstract the mathematical model for XOL. This is listed as the three relations and two function in Table 5. The bracketed types correspond to the higher-order nature of XOL. The slot type within the bracket in the domain of the slot-values relation requires the reification of slots.

**Binary Relations**

```
subclass-of : class → class
instance-of : [class + ] individual → class
slot-values : [class + slot + ] individual → slot × (individual + literal)
```

**Functions**

```
domain : slot → class
slot-value-type : slot → class + datatype
```

**Table 5: XOL Mathematical Model**

From the XOL type hierarchy in Figure 8 and the mathematical model in Table 5 we can identify the correspondences between XOL elements/attributes and the central core with higher-order types. This are listed in Table 6.

| XOL notion | central core notion |
|---|---|
| module, ontology, kb, database, dataset elements | ontology, collection elements |
| class element | Type.Object element |
| name element (within class) | name attribute of object type |
| subclass-of element | subtype element |
| datatype (added type) | Type.Data element |
| class **+** datatype | Type.Entity element |
| slot element | Type.BinaryRelation element |
| name element (within slot) | name attribute of binary relation type |
| domain element | source.Type attribute of binary relation type |
| slot-value-type element | target.Type attribute of binary relation type |
| individual element | Individual.Object element |
| name element (within individual) | id attribute of object instance |
| instance-of element | classification element |
| literal (added type) | Individual.Data element |
| individual **+** literal | Individual.Entity element |
| slot-values element | Individual.BinaryRelation element |
| name element (within slot-values) | type name for binary relation or function |
| value | target.Instance attribute of binary relation instance |
| slot-inverse element | transpose element |
| documentation element | comment element |

**Table 6: Correspondences between XOL and Simple OML**

In Figure 9 places the XOL types in a classification-projection diagram that corresponds to the classification-projection diagram for higher-order types in Figure 4.

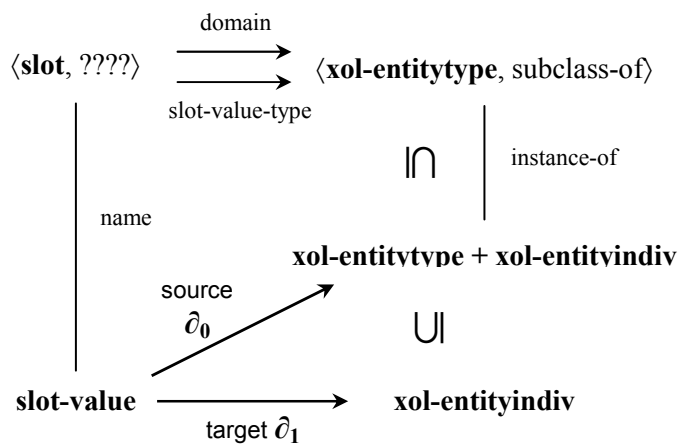

**Figure 9: Classification-Projection Diagram: XOL**

Figures 10 represents interoperability between XOL modules and OML/CKML ontologies and collections in generic style. For interoperability with specific style collections see the discussion on Parsing. The output from the internal representations, and the internal representations themselves, require suitable APIs for XOL and OML.

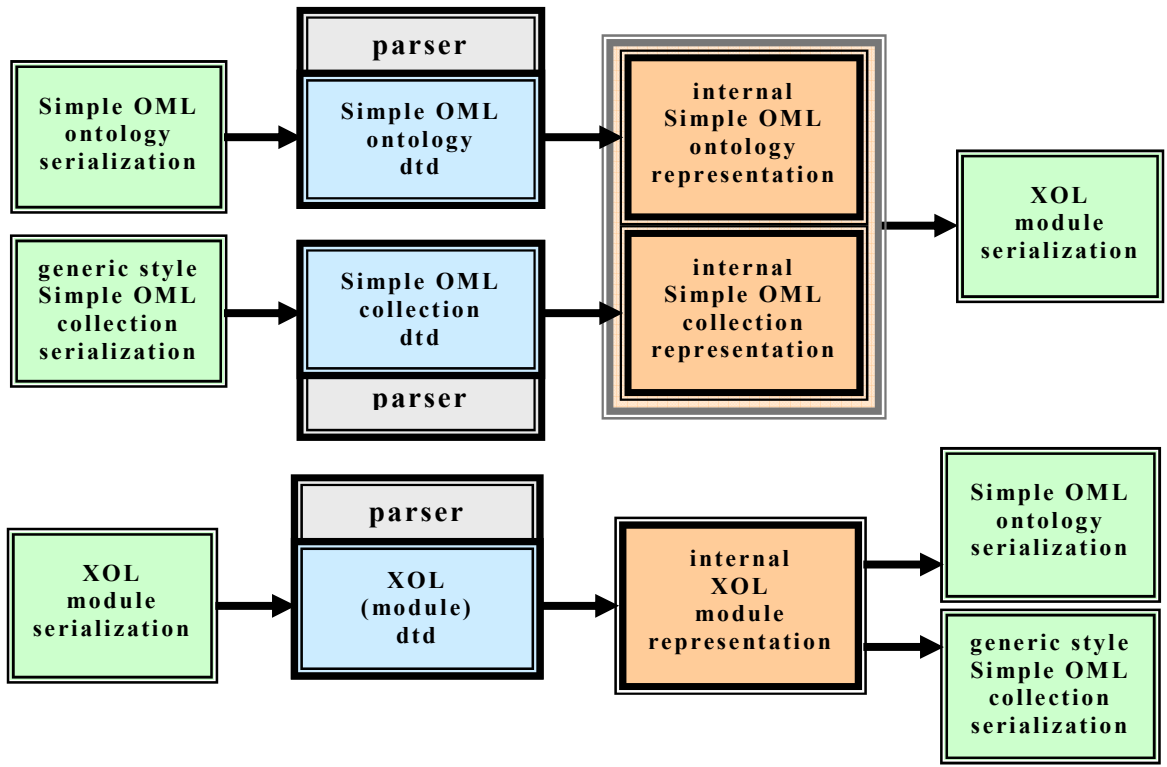

Figure 10: Interoperability between XOL and Simple OML